# Topological phonons in an inhomogeneously strained silicon-5: Inhomogeneous magnetoelectronic effect in a conductor

Paul C. Lou[1], Ravindra G. Bhardwaj[1], Anand Katailiha[1], Ward Beyermann[2], and Sandeep Kumar[1,*]

[1] Department of Mechanical Engineering, University of California, Riverside, CA 92521, USA

[2] Department of Physics and Astronomy, University of California, Riverside, CA 92521, USA

[*] Corresponding author

Email: sandeep.suk191@gmail.com

Abstract

The spatially inhomogeneity in a magnetic crystal give rise to electric polarization, which is known as inhomogeneous magnetoelectric effect. Similarly, an inhomogeneous magnetoelectronic effect in a conducting multiferroic material give rise to spatially inhomogeneous magnetic moment and spin distribution due to spatially inhomogeneity in the charge carrier concentration. In this study, we present experimental evidence of inhomogeneous magnetoelectronic effect in Py/p-Si layered structure. The Py/p-Si layered structure exhibit electronic multiferroicity due to superposition of flexoelectronic charge carrier doping and topological phonons. It gives rise to spatially modulations in the spin density and magnetic moment, which are discovered using the Hall effect measurement. The charge carrier density as well as type of the charge carrier are found to be a function of spatial coordinate as well as direction of magnetic field. The observed modulations can also be interpreted as incommensurate SDW with wavelength of ~142 µm. The inhomogeneous magnetoelectronic effect also give rise to magnetocaloric effect, which is uncovered using thermal hysteresis in the magnetoresistance measurement. This is a first experimental evidence of inhomogeneous magnetoelectronic effect, which is electronic counterpart of the magnetoelectric effect.

The conventional magnetoelectric multiferroic materials[1,2] exhibit simultaneous magnetic and ferroelectric ordering that allow control of the spin and the charge in electric and magnetic fields[3], respectively. However, the magnetoelectric multiferroic materials are insulators and may not be useful in all applications. Recently, the flexoelectronic charge carrier doping of a degenerately doping of the Si thin film in a metal/Si bilayer structure due to strain gradient was demonstrated, which give rise to excess charge carrier concentration in the impurity band of the Si layer. The superposition of topological phonons[4,5] in an inhomogeneously strained Si and flexoelectronic charge doping in the Si thin films give rise to topological electronic magnetism of phonons[6] as shown in Figure 1 (a) and can be described as[7]:

$$\boldsymbol{M}_t \propto \boldsymbol{P}_{FE} \times \partial_t \boldsymbol{P} \qquad (1)$$

$$\boldsymbol{M}_t \propto \frac{\partial n}{\partial z} \times f(\boldsymbol{A}, \boldsymbol{p}) \qquad (2)$$

where $\boldsymbol{M}_t$ is temporal magnetic moment, $\boldsymbol{P}_{FE} \propto \frac{\partial \epsilon_{xx}}{\partial z} \propto \frac{\partial n}{\partial z}$ is flexoelectronic effect due to strain gradient $\left(\frac{\partial \epsilon_{xx}}{\partial z}\right)$ that give rise to gradient of charge carrier concentration $\left(\frac{\partial n}{\partial z}\right)$ and $\partial_t \boldsymbol{P} \propto f(\boldsymbol{A}, \boldsymbol{p})$ is time evolution of topological phonon polarization that is a function of Berry gauge potential $\boldsymbol{A}$ and momentum $\boldsymbol{p}$. We have reported experimental evidence of the topological electronic magnetism in part 3 and 4 of this series. From equation 2, we observed that the momentum, charge carrier density gradient (flexoelectronic polarization) and temporal magnetic moment are coupled to each other. Hence, the behavior described in the equation 2 can also be considered conducting magnetoelectronic multiferroic, which can give rise to multiferroic behavior in a metallic system. Here, instead of ferroelectric polarization and magnetic moment, the charge carrier density and temporal magnetic moment of topological phonons are coupled. Additionally, we posit that the topological phonons having temporal magnetic moment can be considered as an electromagnon in a multiferroic material as shown in the Figure 1 (a). The electromagnon will give rise to spatial modulations in the charge carrier density in this case. This behavior is similar to the inhomogeneous magnetoelectric effect[8] that arises due to the spatial magnetic inhomogeneity in a magnetic crystal[9,10]. Additionally, the topological electronic magnetism will lead to spin splitting of the charge carrier density. As a consequence, the superposition of the charge carrier density and time dependent topological magnetic moment will also lead to spatially modulated spin density fluctuations, which can be experimentally determined.

In this study, we demonstrated an experimental evidence of the spatial modulations in the spin dependent charge carrier density and magnetic moment, which are deduced from the spatial inhomogeneity in the Hall resistance and anomalous Hall effect responses. The spatial inhomogeneity of charge carrier concentration and magnetic moment was expected to arise due to inhomogeneous magnetoelectronic effect. This was the first experimental evidence of the conducting multiferroic material.

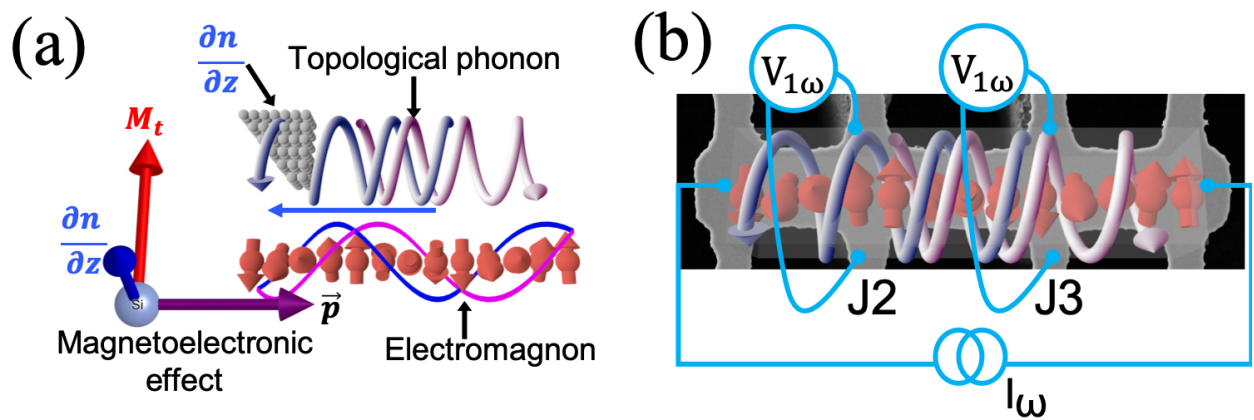

Figure 1. (a) A schematic showing the mechanistic origin of topological electronic magnetism, magnetoelectronic effect and resulting electromagnon, and (b) a schematic showing the measurement scheme for the spatially modulating spin density and a representative sample structure.

To measure the spatially modulated spin density distribution, we used a sample structure with Hall bars. Since our sample was metallic, we used a strategy of Hall effect measurement at two different locations to verify the existence of spatial modulations of spin density as shown in Figure 1 (b). We fabricated a Py/MgO/p-Si (400 nm) Hall bar device[11] having structure similar to previous studies (part 1,2 and 3) as shown in Figure 1 (b).

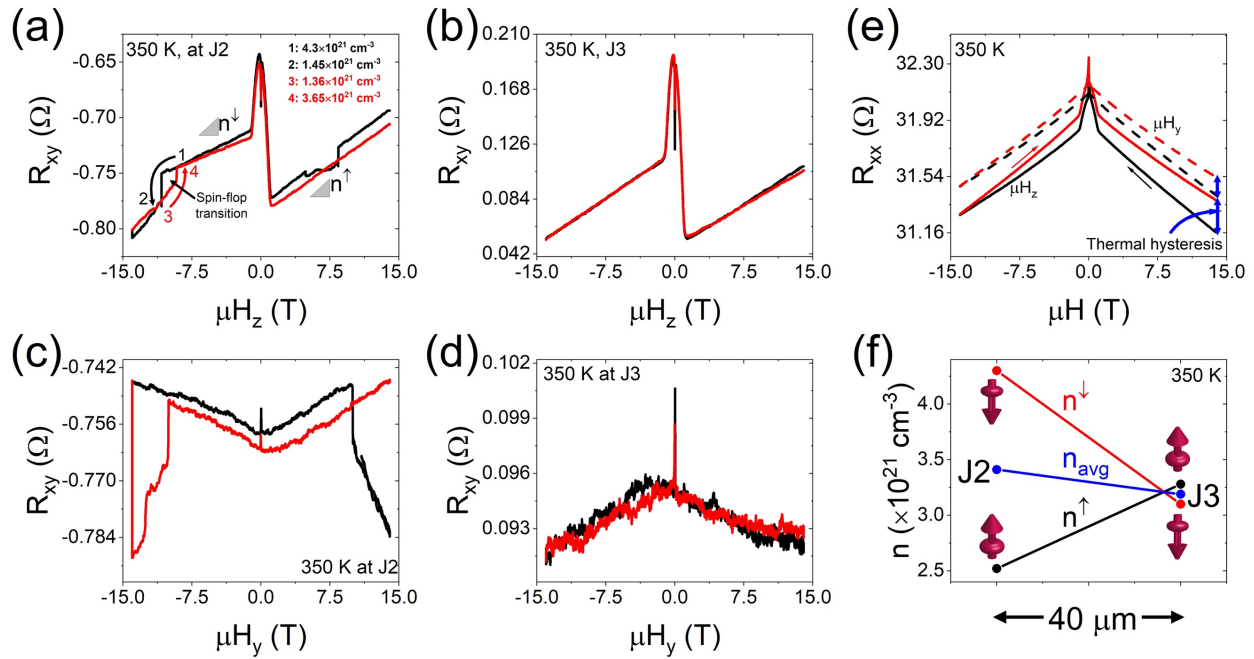

Figure 2. The transverse resistance measurement measured at 350 K and (a) at J2 and (b) at J3 for an out of plane magnetic field, (c) at J2 and (d) at J3 for in-plane (y-dir) magnetic field. (e) The longitudinal magnetoresistance as a function of in-plane and out

of plane magnetic field from 14 T to -14 T at 350 K, and (f) the spatial modulation in the up-spin, down-spin and average charge carrier density.

First, we measured the Hall resistance at 350 K as a function of magnetic field from 14 T to -14 T at two Hall junctions J2 and J3 as shown in Figure 2 (a,b), respectively. The Hall responses measured at J2 and J3 were different from each other but slope was positive in both cases. It meant that the dominant charge carriers were holes. The flexoelectronic charge transfer from Py to Si led to electron deficiency in the Py layer, which was the underlying cause of the positive Hall resistance. In the Hall measurement at J2, we also observed a hysteretic response as shown in Figure 2 (a). This hysteretic behavior was absent in the measurement at J3 even though these measurements were undertaken simultaneously. The hysteretic behavior could be interpreted as a spin flop transition in the local antiferromagnetic superposition from topological phonons. However, the Hall resistance changed due to hysteretic behavior. Based on Hall resistances, the charge carrier concentration reduced from $4.3 \times 10^{21}$ cm$^{-3}$ to $1.45 \times 10^{21}$ cm$^{-3}$ at the magnetic field -10.8 T as shown in Figure 2 (a). In the inverse magnetic sweep, the charge carrier concentration increased from $1.36 \times 10^{21}$ cm$^{-3}$ to $3.65 \times 10^{21}$ cm$^{-3}$ at the magnetic field 9.2 T as shown in Figure 2 (a). Hence, the coercivity of the local magnetic moment was estimated to be 0.8 T and exchange bias of 10 T. The large magnetic field led to spin of charge carrier aligning with the external magnetic field. As a consequence, the reduction in charge carrier concentration was attributed to the transfer of holes above the Fermi level and became electrons. The hysteretic behavior was not observed in the Hall measurement at J3, which clearly showed that local spatial modulations were expected to give rise to the hysteretic behavior. This was the manifestation of the inhomogeneous magnetoelectronic effect since magnetic field led to change in local charge carrier concentration and Hall resistances. We also measured the transverse resistance as a function of in-plane magnetic field (y-direction) as shown in Figure 2 (c,d). The in-plane transverse responses arose due to topological spin Hall effect of phonons and resulting skew scattering. It can also be called as topological planar Hall effect of phonons. Similar to the Hall resistance at J2, the transverse resistance also exhibited hysteretic behavior as shown in Figure 2 (c). The transverse resistance for in-plane magnetic field measured at J3 was qualitatively different from the response measured at J2 as shown in Figure 2 (d) and (c), respectively.

We, then, measured the longitudinal magnetoresistance for in-plane and out of plane magnetic field from 14 T to -14 T. The magnetoresistance response was expected to arise from Py layer only and anisotropic magnetoresistance behavior was observed as shown in Figure 2 (e). However, the resistance at out of plane magnetic field of 14 T increased from 31.165 $\Omega$ to 31.38 $\Omega$ after magnetic field was swept from 14 T to -14 T and back to 14 T. Similarly, the resistance at in-plane magnetic field of 14 T increased from 31.408 $\Omega$ to 31.53 $\Omega$ after magnetic field was swept from 14 T to -14 T and back to 14 T. It is to be noted that instrumental thermal drift lead to decrease in resistance as opposed to the increased resistance observed in this measurement. In addition, the Hall effect responses were measured together with the longitudinal magnetoresistance and no thermal hysteresis was observed. As described earlier, we have observed a magnetic hysteretic behavior in the Hall resistance measurement as shown in Figure 2 (a). However, similar magnetic hysteretic behavior might arise at multiple locations along the length of the sample, which generated heat due to spin dependent electron-phonon scattering. As a consequence, the temperature as well as resistance of the sample increased, which gave rise to thermal hysteresis. Hence, this increase in the resistance due to magnetic field cycling was attributed to the thermal hysteresis from inhomogeneous magnetoelectronic effect and the resulting thermal effect could be interpreted as magnetocaloric effect from spin-phonon interactions.

From the slope of Hall resistance for positive magnetic field, we estimated the charge carrier concentration for spin-up charge carriers and vice-versa for spin-down charge carriers. We found that the concentration of the spin-down charge carrier decreased from left (J2) to right (J3) whereas concentration of spin-up charge carriers increased from left to right as shown in Figure 2 (f). In addition, the average charge carrier concentration also decreased from left to right as shown in Figure 2 (f). This variation in the spin dependent charge carrier concentration was attributed to the spin density modulation, which can be considered as an incommensurate SDW due to superposition of electromagnons. The observation of the spatial spin density modulations was also a proof of inhomogeneous magnetoelectronic effect. We observed a cross over between spin-up and spin-down charge carrier concentration at a distance of ~35.5 µm from the J2 junction. Assuming the distance to be a quarter of wave, we estimated the wavelength of incommensurate SDW to be ~142 µm.

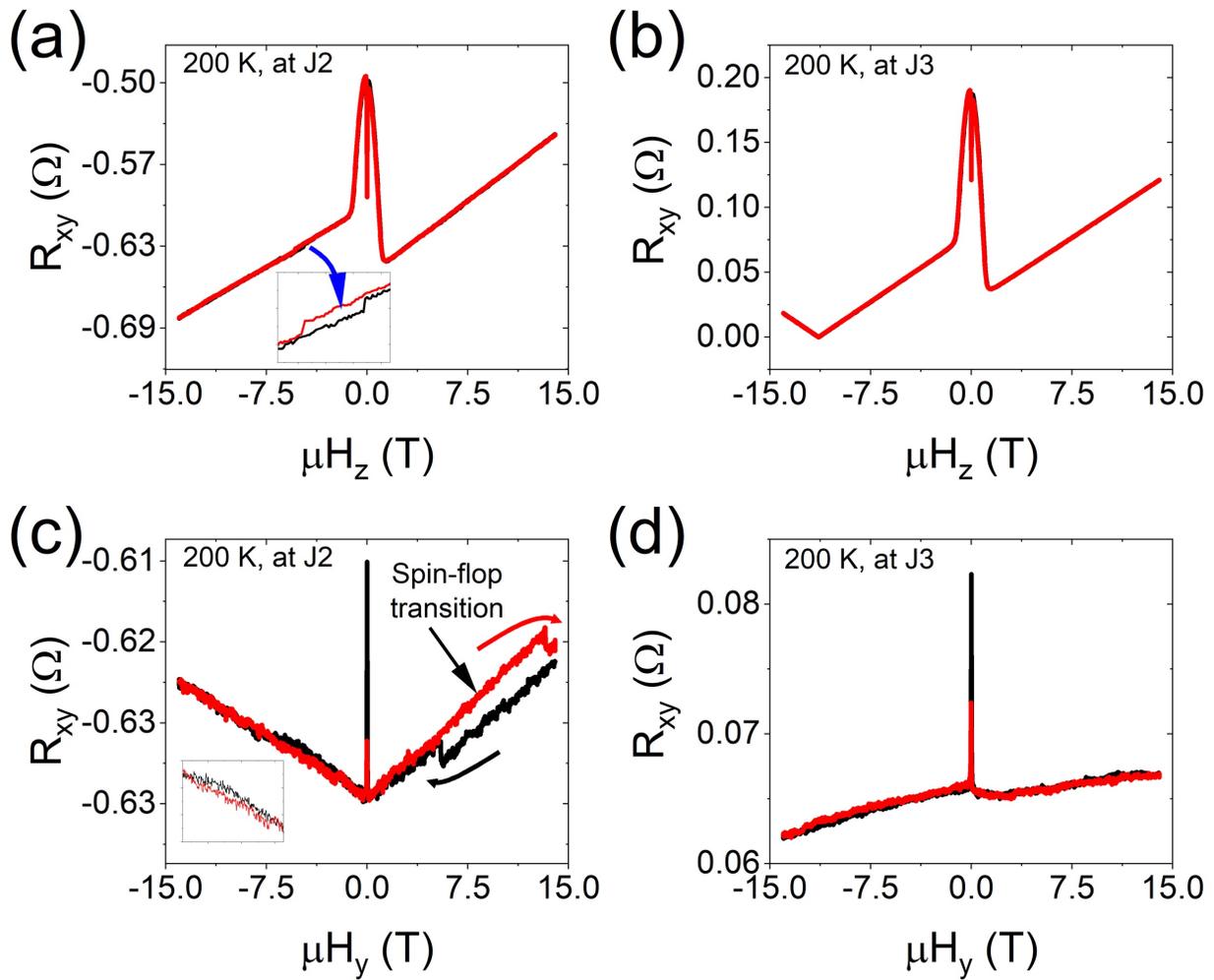

Figure 3. The transverse resistance measurement measured at 200 K and (a) at J2 and (b) at J3 for an out of plane magnetic field, (c) at J2 and (d) at J3 for in-plane (y-dir) magnetic field.

We, then, measured the Hall responses at 200 K for an out of plane magnetic field as shown in Figure 3 (a,b). The spin-flop transition was at lower magnetic field and weaker in magnitude as compared to 350 K as shown in Figure 3 (a) inset. This was attributed to freezing of the topological phonons and resulting decrease in the temporal magnetic moment. In the Hall measurement at J3, we observed a sharp slope change with no hysteresis indicating a transition from hole to electron mediated transport. In addition, the slope on both sides of transition was exactly same, which meant that the charge carrier concentration was same as well. The sign reversal was attributed to the change in Fermi level due to applied magnetic field. Hence, this sign reversal of the Hall resistance was also a manifestation of the inhomogeneous magnetoelectronic effect.

We also measured the transverse response for in-plane magnetic field. Which would eliminate the Hall effect and anomalous Hall effect response of the Py layer. The measured responses at J2 and J3 are shown in Figure 3 (c,d). The transverse resistances behavior at J2 and J3 were qualitatively different from each other. This difference was attributed to the local spin density as well as magnetic moment of topological phonons. In the transverse resistance response measured at J2, we observed a magnetic hysteresis behavior for both positive and negative magnetic fields as shown in Figure 3 (c). This response was also attributed to the spin-flop transition. At 350 K, the topological phonons had magnetic moment aligned out of plane whereas at 200 K it was in-plane, which gave rise to magnetic hysteresis observed in these measurements. However, the magnetoresistance measurements did not exhibit any thermal hysteresis as shown in Figure 4 (a) at 200 K as compared to 350 K. This behavior was attributed to reduction the spin dependent electron-phonon scattering due to freezing of topological phonons.

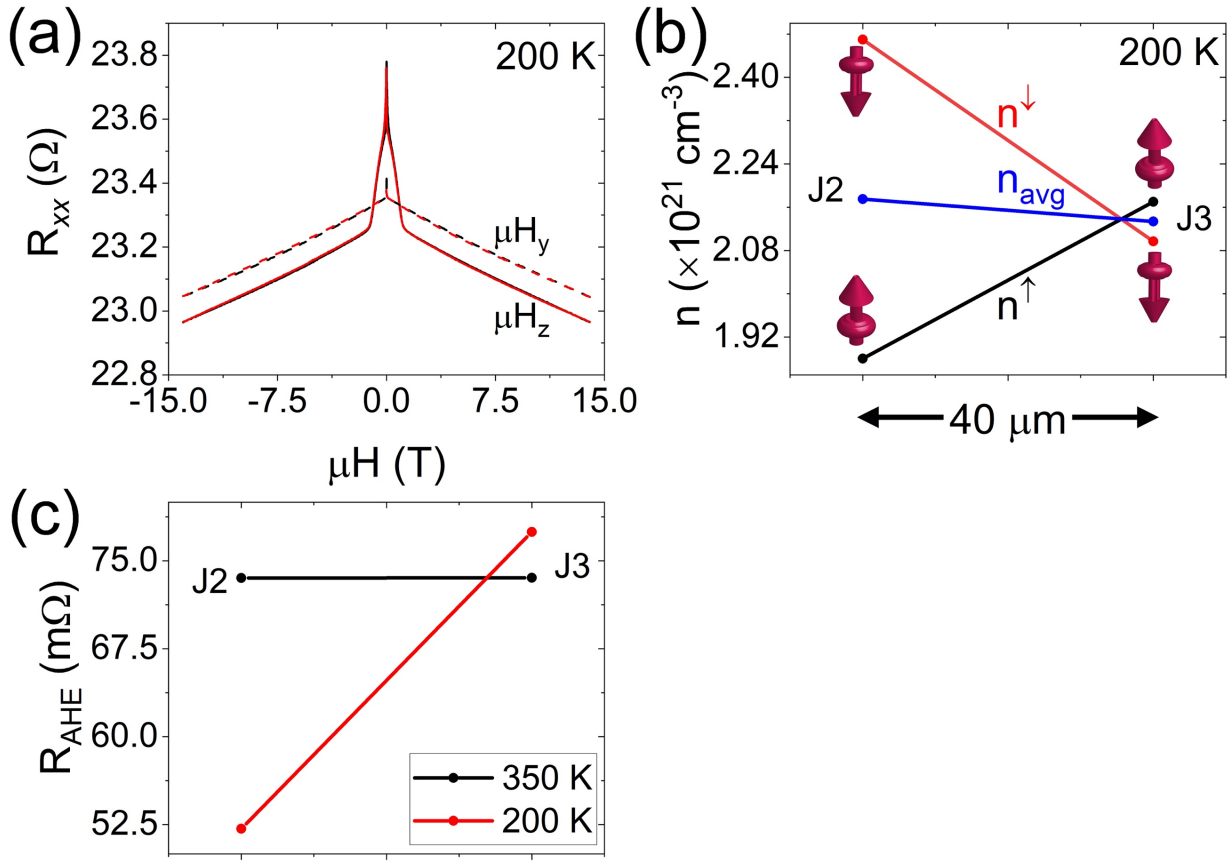

Figure 4. (a) The longitudinal magnetoresistance as a function of in-plane and out of plane magnetic field from 14 T to -14 T at 200 K, (b) the spatial modulation in the up-spin, down-spin and average charge carrier density, and (c) the longitudinal modulation in the anomalous Hall resistance measured at 350 K and 200 K.

We, then, calculated the concentration of spin-up and spin-down charge carrier using Hall effect measurement as shown in Figure 4 (b). We observed positive and negative slope for spin-up and spin-down charge carrier as shown in Figure 4 (b), respectively, which was a clear indication of spin density fluctuations similar to the measurement at 350 K. In addition, the wavelength of the spin modulations or incommensurate SDW to be ~142 µm similar to measurement at 350 K. We, then, estimated the anomalous Hall resistances for all the measurements. At 350 K, the magnitude and sign of the anomalous Hall resistance (73.55 mΩ) was same at both J2 and J3 as shown in Figure 4 (c). However, at 200 K, the anomalous Hall resistance was significantly smaller at J2 (52.15 mΩ) as compared to J3 (77.46 mΩ). This reduction in the anomalous Hall resistance was attributed to the magnetic moment of topological phonons aligning antiferromagnetically to the magnetic moment of Py layer. This measurement again demonstrated longitudinal spin density modulation due to the inhomogeneous magnetoelectronic effect.

In conclusion, we have demonstrated inhomogeneous magnetoelectronic effect in conducting Py/p-Si layered structure. The Py/p-Si layered structure exhibit electronic multiferroicity due to superposition of flexoelectronic charge carrier doping and topological phonons. We observed spatial modulations in the spin density and magnetic moment in the Hall effect measurement. The charge carrier density as well as type of the charge carrier are found to be a function of spatial coordinate as well as direction of magnetic field. The inhomogeneous magnetoelectronic effect also give rise to magnetocaloric effect, which was uncovered using thermal hysteresis in the magnetoresistance measurement. This is a first experimental evidence of inhomogeneous magnetoelectronic effect, which is electronic counterpart of the magnetoelectric effect. This discovery has potential to revolutionize the multiferroic materials research and applications.

**Acknowledgement**

The fabrication of experimental devices was completed at the Center for Nanoscale Science and Engineering at UC Riverside. Electron microscopy imaging was performed at the Central Facility for Advanced Microscopy and Microanalysis at UC Riverside. SK acknowledges a research gift from Dr. Sandeep Kumar.